\documentclass[a4paper,11pt]{article}
\usepackage[utf8]{inputenc}
\usepackage{authblk}
\usepackage{setspace}
\usepackage{graphicx}
\usepackage{subcaption}
\usepackage{amsmath}
\usepackage[margin=1.0in]{geometry}
\usepackage{url,hyperref,lineno,multirow,graphicx}
\usepackage{textcomp}
\usepackage[numbers,square,sort&compress]{natbib}
\title{ \textbf{Future Developments in Charged Particle Therapy: Improving Beam Delivery for Efficiency and Efficacy} }

\author[1*]{Jacinta Yap}
\author[2]{Andrea De Franco}
\author[1]{Suzie Sheehy}
\affil[1]{School of Physics, University of Melbourne, Melbourne, VIC, Australia.}
\affil[2]{National Institute for Quantum and Radiological Science and Technology (QST), Rokkasho-mura, Aomori, Japan.}

\affil[*]{\small{Corresponding author. Email: jacinta.yap@unimelb.edu.au}}

\date{}
\begin{document}

\maketitle


\begin{abstract}
The physical and clinical benefits of charged particle therapy (CPT) are well recognised and recent developments have led to the rapid emergence of facilities, resulting in wider adoption worldwide. Nonetheless, the availability of CPT and complete exploitation of dosimetric advantages are still limited by high facility costs and technological challenges. There are extensive ongoing efforts to improve upon these, which will lead to greater accessibility, superior delivery, and therefore better treatment outcomes. The issue of cost however still remains a primary hurdle as the utility of CPT is largely driven by the affordability, complexity and performance of current technology. Several of these aspects can be addressed by developments to the beam delivery system (BDS) which determine the overall shaping and timing capabilities to provide high quality treatments. Modern delivery techniques are necessary but are limited by extended treatment times. The energy layer switching time (ELST) is a limiting constraint of the BDS and a determinant of the beam delivery time (BDT), along with the accelerator and other factors. This review evaluates the delivery process in detail, presenting the limitations and developments for the BDS and related accelerator technology, toward decreasing the BDT. As extended BDT impacts motion and has dosimetric implications for treatment, we discuss avenues to minimise the ELST and overview the clinical benefits and feasibility of a large energy acceptance BDS. These developments support the possibility of delivering advanced methodologies such as volumetric rescanning, FLASH and arc therapy and can further reduce costs given a faster delivery for a greater range of treatment indications. Further work to realise multi-ion, image guided and adaptive therapies is also discussed. In this review we examine how increased treatment efficiency and efficacy could be achieved with an improved BDS and how this could lead to faster and higher quality treatments for the future of CPT. 
\end{abstract}

\section*{}
\small 
\textbf{Keywords:} \\
\\
\indent particle therapy, beam delivery, accelerators, large energy acceptance, energy layer switching time, rescanning, FLASH, Arc therapy

\newpage
\section{Introduction}
Access to charged particle therapy is growing rapidly worldwide. As a therapeutic modality CPT is now well established, where proton beam therapy (PBT) is the most common type. Growth in the number of facilities has been bolstered by developments in accelerators and related technologies, beam delivery methods, verification tools, and increased clinical experience. Where available, PBT is often standard practice, particularly for paediatric cases and specific tumour types (ocular, head and neck tumours \cite{Engelsman2013}). CPT has an important prospective role in reducing the growing cancer burden on a global scale, and its impact could be significant \cite{Bortfeld2020} however, its full potential is yet to be realised. Overcoming this requires improvements in two key areas: improving efficacy and decreasing cost.\\

CPT offers benefits over and above regular radiotherapy (RT) for palliative or curative treatments, offering not just physical dose escalation but also biological advantage. Yet in terms of efficacy, we cannot capitalise fully on increased radiobiological effectiveness of charged particles at present, primarily due to limitations in knowledge \cite{Schaue2015}. For this reason, existing programmes of research are investigating the underlying mechanisms of CPT in terms of fundamental chemical, biological and cellular processes \cite{Luhr2018,Held2016,Tinganelli2020,Lomax2013,Durante2014,Vitti2019} to try to understand the roles of these processes in determining clinical outcomes \cite{Durante2018a,Durante2017,Rackwitz2019}. Nonetheless, the superior physical properties of charged particles are evident as the characteristic `\textit{Bragg Peak}' (BP) enables a precise dose distribution and an improved therapeutic window.\\

The second reason we cannot yet fully exploit the efficacy of CPT is due to technological limitations. Advanced techniques and technological improvements for CPT seek to deliver higher quality treatments with increased conformity as these translate to long-term benefits. However some of these improvements would increase, rather than decrease, the cost of the treatment.\\

In terms of cost (or efficiency) the gap between conventional X-ray photon RT (XRT) and CPT still exists due to the many challenges to be addressed: affordability, complexity and limitations with current technology all restrict the utility of CPT. Although availability has surged in recent years as several vendors offer competitive and commercial turn-key solutions, high capital and operational costs are still a primary issue. Many potential areas of improvement have been well identified \cite{Paganetti2021,Schreuder2020,Farr2018,Schippers2018,Myers2019,Owen2016,Lomax2018,Flanz2013}. In general, the potential for cost reduction can be considered by decreasing the facility and machine size, operational complexities, treatment times, increasing the treatment workflow efficiency and hence throughput. \\

However, just making the treatment cheap and widespread is not enough. Both efficiency and efficacy need to be improved, in other words, even in an ideal world of low-cost and widespread availability of facilities, the maximum possible clinical benefit of CPT will only be achievable if existing technical limitations are overcome. While there are many factors which govern the overall cost and efficacy of treatment: one key underlying aspect is the beam delivery. Complimentary to the accelerator complex, the beam delivery system is an essential piece in the treatment workflow which contributes significantly to the overall treatment time, and which will form the primary focus of this review.\\

At present, the majority of CPT treatments use active pencil beam scanning (PBS), involving the intricate delivery of several thousand overlapping narrow beams which results in highly conformal dose distributions. Most clinical indications are treated with PBS however an unavoidable and consequential issue is the lengthy beam delivery time. During treatment the beam is scanned across the (potentially large) target site, and the accelerator complex, control system and diagnostic instruments need to adjust. A key bottleneck in this process is the energy layer switching time.\\

The slow ELST is due to technical constraints and is a prevalent issue: it is the critical component which accounts for the majority of the BDT \cite{Shen2017,Suzuki2016}. The beam moves relatively quickly across the tumour transversely, but it takes much longer to switch the beam in depth. This extends the beam-on time and long overall irradiation times can increase dose uncertainties. This is a key problem for cases where the tumour site itself may also move, for example, interplay effects are unavoidable during lung treatments as caused by respiratory motion. Although different motion management approaches such as gating, rescanning and tracking can be performed, the clinical implications reduce the utility of CPT, especially in particular indications \cite{Durante2017}. Moreover, as the high dose BP region is a motivation for using CPT, sensitivity to changes in range are especially important -- and for heavy ions this is even greater -- which impairs benefits attained from the physical and radiobiological advantages \cite{Loeffler2013,Durante2016}. In other words, the beam delivery impacts both efficiency (cost) and efficacy, and is thus a crucial aspect to investigate for the future of CPT.\\

Of course, there are many aspects of CPT, both technological and systems-based which could improve efficacy and cost in future: the scope of this review is to look at the technology pertaining to the BDS, broadly covering those aspects of the system which impact the dose delivery process and treatment time\footnote{Existing recommendations and requirements for technological improvements have been primarily focused on PBT, which precede developments for CPT with heavier ions due to greater clinical experience. We include many references to PBT for readers to explore these where relevant.}. An improved BDS could lead to better treatment efficiency and efficacy, both of which are necessary for advanced delivery methodologies.\\

There are several limitations and central issues surrounding existing BDS capabilities which need to be addressed to improve the delivery of CPT in future. In the near term, can we improve the capabilities of the BDS to better deliver CPT? Can we decrease the BDT? What are the current constraints and where are improvements needed? How does this translate to the clinic? In the longer term, what does the future of CPT delivery look like? How will new treatment paradigms such as FLASH or multi-ion therapy be limited by BDS technology?\\

We review the limitations and developments needed for the BDS and related accelerator technology with the outlook of decreasing the BDT. The clinical feasibility, impact on the delivery process and potential benefits of different approaches are examined, with a focus on the perspective of minimising the ELST, its contribution to the BDT and its implication for treatment.

\section{Beam Delivery}
\label{Section_BeamDelivery}
\begin{figure}[htb!]
    \centering
    \includegraphics*[width=\textwidth]{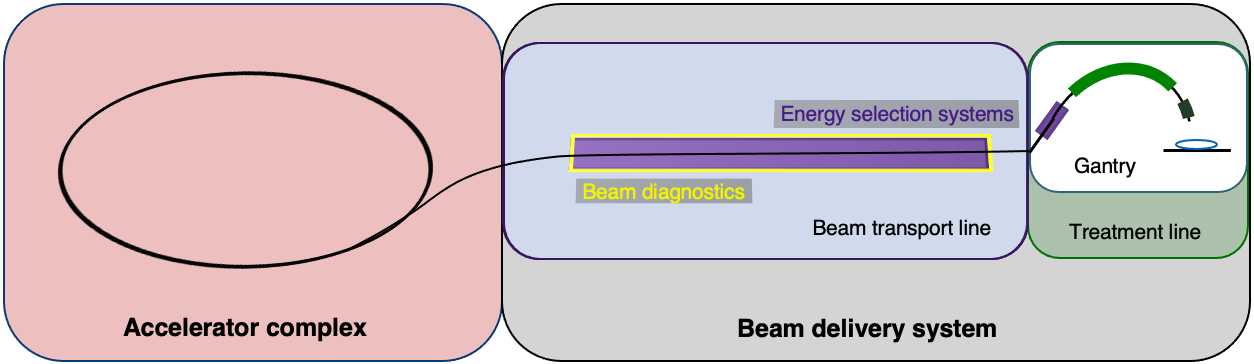}
    \caption{Illustration overviewing the main components in a CPT treatment facility. The BDS is shown to include the BTL (beam instrumentation and devices which may alter the energy or shape of the beam) and treatment line (fixed line or here, a gantry).}
    \label{F_BDScomponentsOverview}
\end{figure}

The beam produced by the accelerator must be shaped and modified to deliver dose to the target site matching the requirements of the treatment plan. This requires changing the distribution of the spatial (lateral) and energy (longitudinal) spread and also often the time structure, i.e. beam modification in four-dimensions is required. These can be tuned to an extent by the accelerator complex and beam transport lines (BTL). In this review we define the BDS as the components after the accelerator which determine how the beam is shaped, transported and ultimately delivered to the patient for treatment. This encompasses the BTL, diagnostic instrumentation, energy selection systems, treatment line (gantry or fixed delivery line) and treatment head. For brevity, we focus on the BTL, treatment line and delivery system for PBS (Figure \ref{F_BDScomponentsOverview}). \\

Treatments can take anywhere from a few minutes to more than half an hour, depending on the tumour type and complexity of the plan. The overall treatment time is determined by the BDT, additional activities including patient setup (immobilisation, imaging, unload etc.) and equipment related checks (couch positioning, beam checks, readying beam devices etc.). The BDT consists of the actual irradiation `\textit{beam-on}' time and the `\textit{beam-off}' time, spent requesting and waiting on the beam after adjustments or between fields. A quantitative analysis of PBT treatments by Suzuki et al. \cite{Suzuki2011} reported that approximately 80\% of treatment time was spent on these additional activities with the remaining 20\% contributed by the BDT. Total treatment time is shown to increase quadratically with the number of fields, with complex cases requiring the same amount of time to carry out patient-related activities but accruing larger contributions from equipment checks and BDT. Reductions in these latter aspects have more potential to improve the efficiency as patient-related process times can vary widely, are circumstantial, depend on the physical and clinical condition of the patient and cannot necessarily be improved with technology. Furthermore, shorter treatment times are preferred not just for cost but also due to difficulty of immobilisation and set-up of patients for 30 minutes or longer. As discussed by Nystrom \& Paganetti et al. \cite{Paganetti2021}, a faster BDT can result in a significant gain in treatment efficiency, particularly for multi-room facilities with high waiting times. Evidently, any increase in treatment efficiency is valuable.\\

\begin{figure}[h!]
    \centering
    \includegraphics*[width=0.78\textwidth]{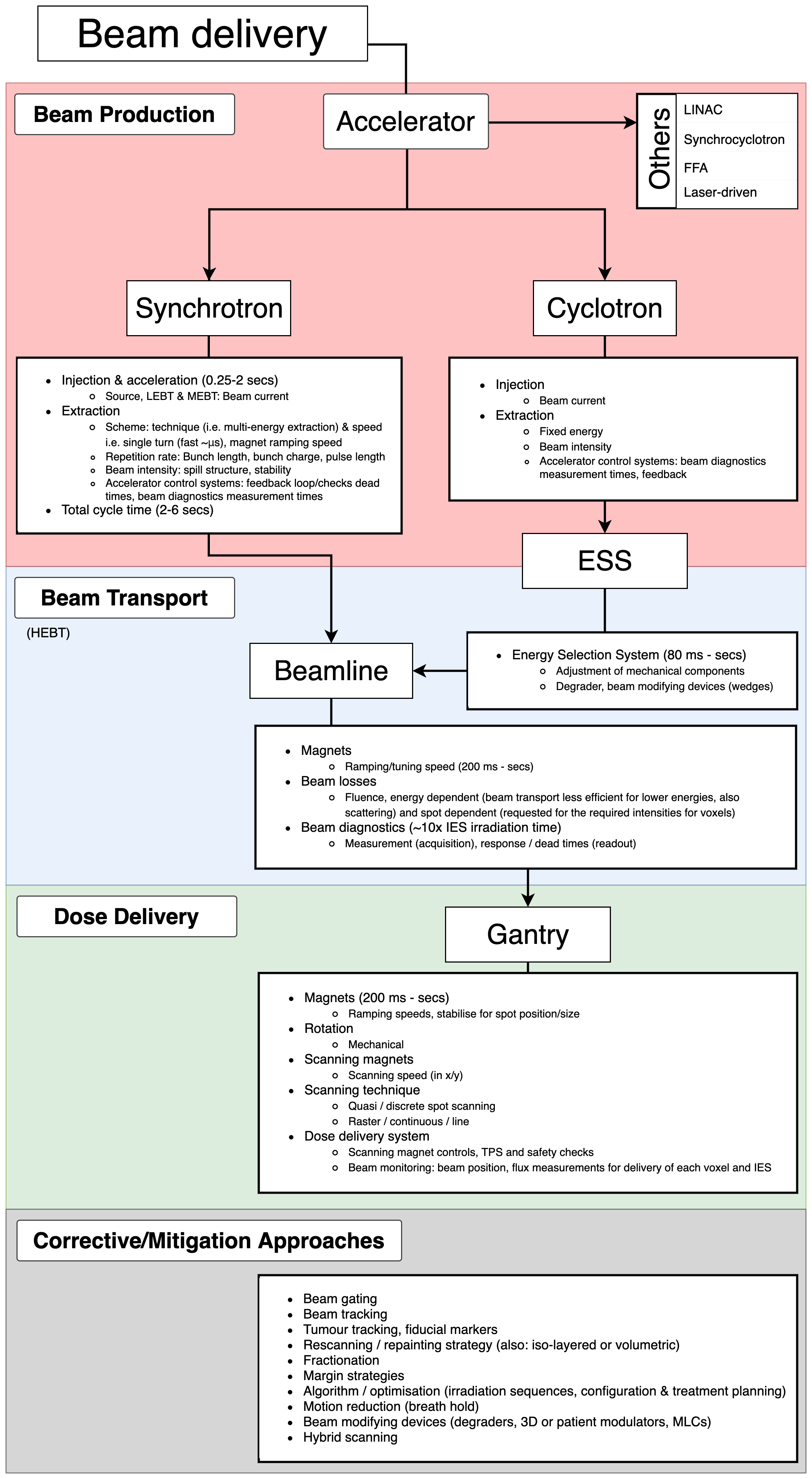}
    \caption{Overview of PBS beam delivery and different motion mitigation strategies. Breakdown of contributions from the beam production, beam transport and delivery processes to BDT.}
    \label{F_DelivChart}
\end{figure}

Decreasing BDT is complex as facilities are not standardised, and neither are prescribed treatment plans. The BDS, accelerator and other systems vary at each facility and the delivery efficiency depends on numerous technology-related factors. Facilities have different equipment vendors, numbers of rooms, delivery systems etc. and often adopt different processes: these characteristics all influence the delivery procedures implemented \cite{Muller2016}. Meanwhile the treatment plan calculates the number of spots and layers to deliver the required dose distribution to the target volume. Nystrom \& Paganetti et al. \cite{Paganetti2021} state the three main components which constitute the delivery time for a treatment field: time to irradiate a spot, time to move between spots and the time to change beam energy. Speeding up any of these components can shorten BDT however they are not independent variables. Delivering a faster treatment is not straightforward; it is not solely dependent on the capabilities of the BDS itself but is a multi-faceted problem. All the contributing factors and their corresponding time budgets can be examined to assess their impact on dose delivery and the BDT, illustrated in Figure \ref{F_DelivChart}.  

\newpage

\newpage
\subsection{Pencil Beam Scanning}
\begin{figure}[htb!]
    \centering
    \includegraphics*[width=\textwidth]{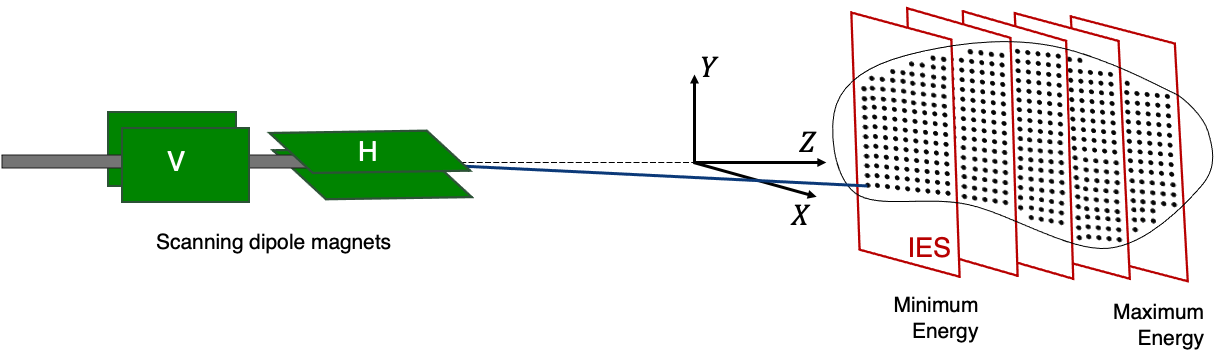}
    \caption{Active PBS delivery. The BDS delivers a conformal dose distribution to the treatment volume by scanning the beam along a calculated path in the transverse plane. The beam energy is then adjusted to change the depth, typically lower in energy, switching to a proximal IES where the subsequent layers are scanned.}
    \label{F_PBS}
\end{figure}

The objective of the BDS is to deliver the beam with the required parameters prescribed by the treatment plan. The planned dose distribution determines the requested parameters and thus system configuration (i.e. together with the source, accelerator, low, medium and high energy beam transport systems). The requested specifications include the beam energies, size at isocentre, intensities and delivery channels: each different configuration can total to thousands of available beam combinations \cite{Lazarev2011}. This results in the delivery of multiple beams which produce a 3D dose distribution and as previously mentioned, the entire process can to amount to long treatment times. For PBS delivery, the beam is magnetically deflected across the tumour in the transverse plane across one layer or an iso-energy slice (IES), then adjusted longitudinally to a shorter depth (typically a decrease in proton range of 5 mm in water) and repeated (Figure \ref{F_PBS}).

\begin{figure}[htb!]
    \centering
    \includegraphics*[width=0.75\textwidth]{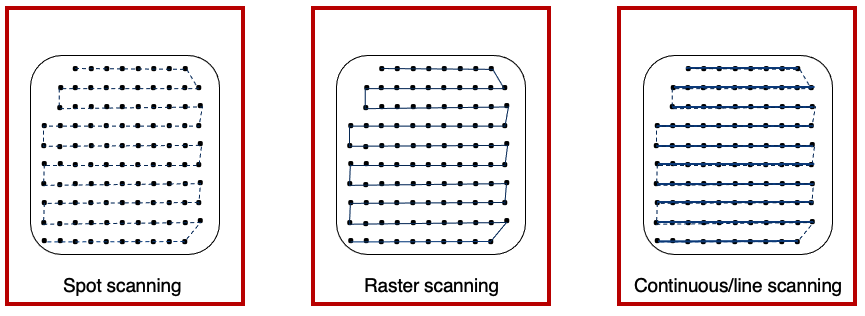}
    \caption{Spot, raster and continuous line scanning patterns for PBS delivery. Spots and solid lines indicate beam-on irradiation and dashed lines indicate movement with beam-off.}
    \label{F_ScanningTech}
\end{figure}

Different scanning techniques (spot, raster and line/continuous) may be used with optimisation methods to deliver the beam and irradiate each layer. The dose is painted such that the accumulation of the distribution in both planes results in sufficient coverage of the tumour volume (Figure \ref{F_ScanningTech}). 

\vspace{4mm}

\begin{figure}[h!]
    \centering
    \includegraphics*[width=0.7\textwidth]{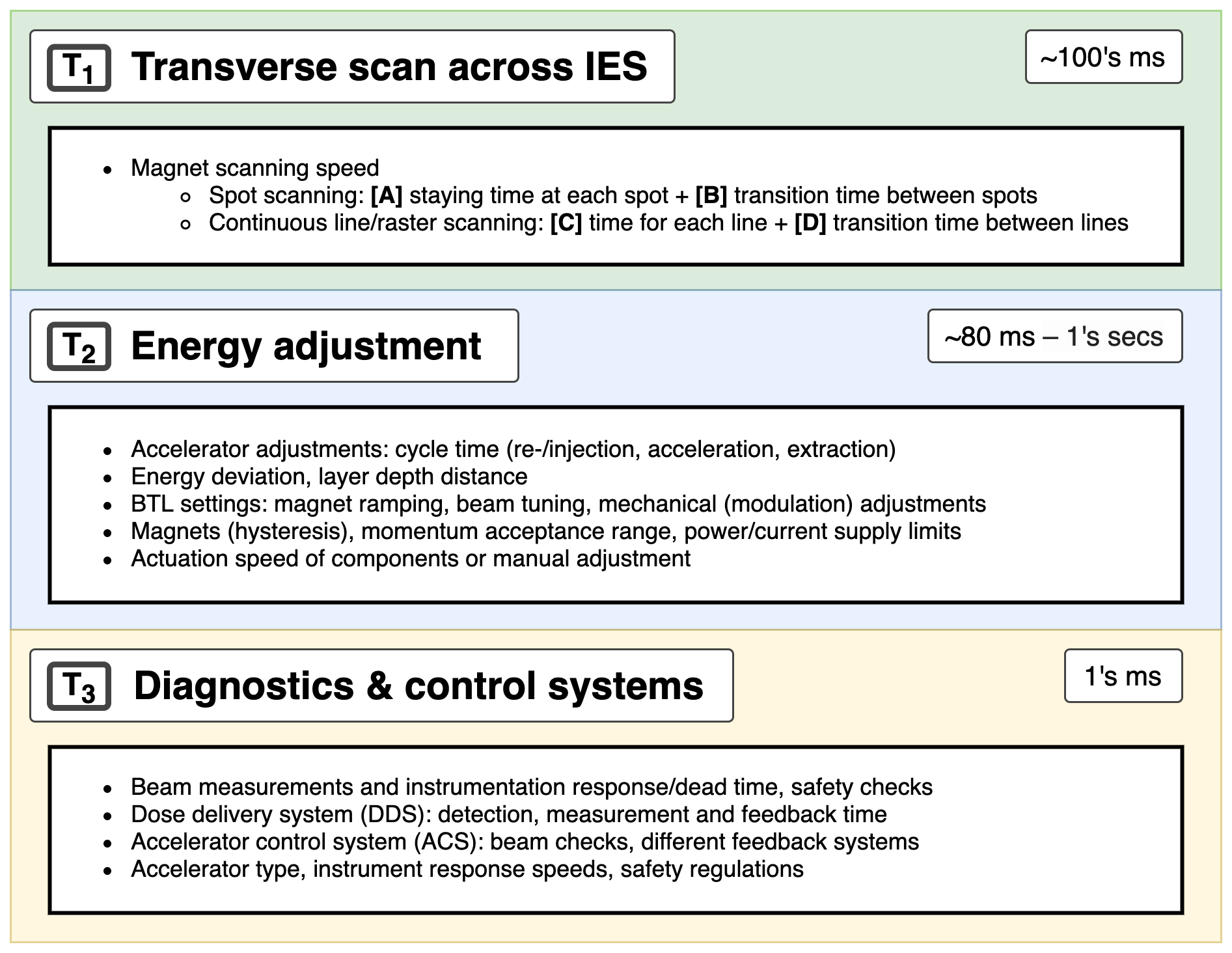}
    \caption{Major BDS components and corresponding factors which contribute to BDT.}
    \label{F_PBS_BDT}
\end{figure}

If we consider only the beam delivery process, this can be approximated to include the \textbf{[T\textsubscript{1}]} transverse scanning, \textbf{[T\textsubscript{2}]} energy adjustment and \textbf{[T\textsubscript{3}]} systematic dead times (Figure \ref{F_PBS_BDT}). 

\newpage
\subsubsection{BDT time components}
\label{Section_BDTtimecomp}
Each IES irradiation engages multiple aspects of the BDS, combining to the total BDT. The time components of a typical treatment can be examined by breaking down the PBS delivery process. As a general example, the BDT can be approximated by summing the different contributions by estimating time allocations associated with each of the 3 factors. \\

\textbf{[T\textsubscript{1A}]} the staying or `\textit{dwell}' time at a position can be down to 0.1 ms and the \textbf{[T\textsubscript{1B}]} transition time to the next position, 0.03 ms \cite{Furukawa2010}. For line scanning, \textbf{[T\textsubscript{2A}]} could be 5 ms for a line and \textbf{[T\textsubscript{2B}]} 5 ms to move to the next line \cite{Pedroni2004}. For an IES, it could take 10's ms altogether for raster scanning \cite{Saito2009}. \textbf{[T\textsubscript{1}]} Irradiation time for a single IES depends on the size of the distribution but needs at least $\sim$100 ms.\\

\textbf{[T\textsubscript{2}]} ranges from 80 ms to a few seconds \cite{Farr2018}, corresponding to faster dynamic modulation with cyclotrons and synchrocyclotrons, or slower direct energy adjustments with synchrotrons. The fastest reported ELSTs are listed in Table \ref{T_AccBaselineFigs}.\\

\textbf{[T\textsubscript{3}]} ionisation chamber measurement times average $\sim$0.1 ms \cite{Lazarev2011,Paganetti2021}. ACS and diagnostic safety checks ensure correct beam parameters (spot size, position, intensity etc.). As reported by Schoemers et al. \cite{Schoemers2015} measurement times are a bottom line limit across the BDS and similarly for continuous scanning, the lower bound is determined by the instrumentation speed \cite{Flanz2013}; at least 1 ms is required. \\

These estimations are provided (within applicable orders of magnitude) as based on broadly reported values and for context, can be expressed simply by: \[BDT = (N)\textbf{T}_\textbf{1}+(N-1)\textbf{T}_\textbf{2} +(N)\textbf{T}_\textbf{3}.\]
Where \textit{N} is the number of IES. For example, a BDT estimate for a simple case with fast timing budgets and a once, single directional irradiation, for 30 IES: 30$\times[$\textbf{T\textsubscript{1}} = 150 ms] + 29$\times$[\textbf{T\textsubscript{2}} = 500 ms] + 30$\times$[\textbf{T\textsubscript{3}} = 2 ms] = 19 s. \\

The BDT is a function of the irradiation sequencing and indeed, a larger tumour volume or higher number of IES recruits more of these actions \textbf{[T\textsubscript{1--3}]}, amounting to a longer BDT. It is noted that the actual beam-on irradiation time \textbf{[T\textsubscript{1A}]}  is very short compared to the beam-off \textbf{[T\textsubscript{2}]}  and dead times \textbf{[T\textsubscript{3}]}; often the total dead time exceeds the irradiation time. Independently, the irradiation time is essentially determined by the intensity of the beam produced by the accelerator \cite{Farr2018,Schoemers2015}. As a standard (PBT) clinical minimum, most facilities have the capability of delivering dose rates of 2 Gy/min to a 1 L volume, 10--20 cm deep \cite{Jolly2020}. This equates to beam currents of 100's nA, varying for accelerator type. However, even at facilities which are able to achieve higher dose rates, there are practical limitations with operation at higher intensities. These depend on machine characteristics as well as safety regulations and instrumentation constraints, as experienced with attempts to reach FLASH rates \cite{VandeWater2019,Nesteruk2021a}. \\

Alternatively, the transverse motion of the beam can be sped up by using continuous scanning methods and with faster dipole magnets. As discussed by Flanz \& Paganetti et al. \cite{Paganetti2021} scanning dipoles from 3--100 Hz are used clinically but capabilities also depend on their size, distance from the patient as well as inductance and power supply considerations. Furthermore, speeds are restricted by the viability of currently available beam instrumentation tools to accurately measure and rapidly record dose rates. 

\subsection{Energy Layer Switching Time}
Decreasing the time delays imposed by \textbf{[T\textsubscript{2}]} changing energy between IES appears to be the more challenging issue. Similarly, it is not just a singular aspect of the delivery process but is governed by the accelerator and significantly, the BDS. The ELST across facilities ranges widely and can be up to an order of magnitude longer than the time it takes to scan across an IES. If we consider the breakdown of time components again (Section \ref{Section_BDTtimecomp}) but this time applying a fast ELST (100 ms) instead for comparison: 30$\times[$\textbf{T\textsubscript{1}} = 150 ms] + 29$\times$[\textbf{T\textsubscript{2}} = 100 ms] + 30$\times$[\textbf{T\textsubscript{3}} = 2 ms] = 6.9 s. \\

\begin{figure}[htb!]
    \centering
    \includegraphics*[width=\textwidth]{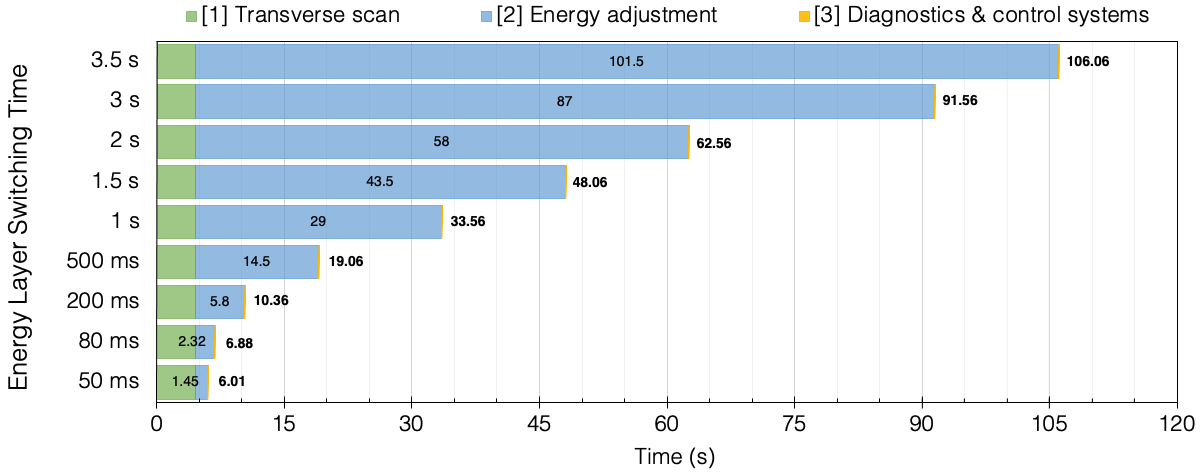}
    \caption{Total BDTs estimated for a range of applicable clinical ELSTs (\textbf{T\textsubscript{2}}) with fixed times for the transverse scan (\textbf{T\textsubscript{2}}) and  system dead times (\textbf{T\textsubscript{3}}), for each IES. Timings are based on previously reported information, where values were made available. For all 30 IES, \textbf{T\textsubscript{1}} = 4.5 s and \textbf{T\textsubscript{3}} = 0.06 s.}
    \label{F_ELSTbarchart}
\end{figure}

Evidently, this results in a much shorter BDT. The ELST appears to be the limiting constraint; particularly for cases with many layers which may also need rescanning, there is an accumulation of time saved at each step. The increasing penalty for longer ELSTs on BDT is illustrated in Figure \ref{F_ELSTbarchart}.

\vspace{4mm}

Several studies have been performed which examine the time components quantitatively and evaluate the impact of the ELST on BDT, as a means to improve treatment delivery efficiency. Shen et al. \cite{Shen2017} carried out a detailed analysis to model the BDT at a synchrotron PBT facility based on operational parameters including the ELST, average scanning speeds, spill rate, charge and extraction time, magnet preparation and verification time. These contributions were similarly reduced to 3 time components (i.e. \textbf{[T\textsubscript{1A}]}, \textbf{[T\textsubscript{1B}]} and \textbf{[T\textsubscript{2}]}) where the average ELST across the energy range and scanning speeds in \textit{x} and \textit{y} were reported as 1.91 s, 5.9 m/s and 19.3 m/s, respectively. Values determined by the model were compared with log files from a large range of delivered patient treatments to calculate the contributions to BDT accurately. The ELST was identified as the most dominant contributor to BDT at 71\%; reducing this time would greatly improve beam utility during delivery. \\

All components of the treatment process were also comprehensively analysed by Suzuki et al. \cite{Suzuki2011} to evaluate the use factor and efficiency of beam delivery parameters for different disease sites. Although there are numerous factors, the BDS largely governs treatment efficiency which is asserted as the most important factor in PBT as it is directly related to utility and availability: for facilities which operate a busy schedule, a 1 min reduction in BDT for a single field could result in the equivalent of treating 10 more (prostate) patients a day. This can also lower costs as treatment costs scale with the total time spent by the patient in the treatment room \cite{Paganetti2021}.\\

Increasing throughput is an important consideration to improve the availability of CPT. A sensitivity analysis of daily throughput capacity – and therefore efficiency of PBS treatments – was subsequently performed at the same facility by Suzuki et al. \cite{Suzuki2016}. Several parameters in the treatment process were similarly studied; the BDT was reduced to the sum of the ELST and spot delivery time as a function of the treatment volume, dependent on the disease site. The ELST was reported as 2.1 s and accounted for 70--90\% of the BDT for the majority of tumour volumes ($<$1 L). Although for this case the BDT is limited by the accelerator, a reduction in the waiting times can greatly decrease BDT: the ELST as well as room switching time account for a large part of the total treatment time. Increasing the uptime by minimising beam-off time can significantly improve the throughput \cite{Aitkenhead2012}. Nystrom and Paganetti et al. \cite{Paganetti2021} emphasise that improvements in this area will have the greatest efficiency gain and that a faster BDT will have the greatest impact on facilities with multiple rooms. Developments in the BDS and accelerator technology are needed for this. 

\begin{table}[b!]
\caption{Reported minimum ELSTs for currently used clinical accelerators: cyclotrons, slow cycling synchrotrons and synchrocyclotrons.}
\label{T_AccBaselineFigs}
\resizebox{\textwidth}{!}{
{\renewcommand{\arraystretch}{1.5}
\begin{tabular}{|l|l|l|l|}
\hline
\multirow{2}{*}{} &
  \multicolumn{3}{c|}{Accelerator type} \\ \cline{2-4} 
    & Cyclotron &
  \multicolumn{1}{c|}{Synchrotron} &
  \multicolumn{1}{c|}{Synchrocyclotron} \\ \hline
\begin{tabular}[c]{@{}l@{}}Energy layer switching times\\ -- fastest reported. \\ \\ Modulation method. \end{tabular} &
  \multicolumn{1}{l|}{\begin{tabular}[c]{@{}l@{}}\textbf{PSI}\\ G2: 80 ms \\ G3: 200 ms \\ \\ ESS, degrader \\ (carbon wedges)\end{tabular}} &
  \begin{tabular}[c]{@{}l@{}}\textbf{HIMAC}\\ MEE: 220 ms \\ Hybrid: 100 ms \\ \\ MEE ($>$3 cm depths) \\ and range shifters ($<$3 cm) \end{tabular} &
  \begin{tabular}[c]{@{}l@{}}\\ \textbf{Mevion S250}\\ 50 ms \\ \\ Motorised modulator \\ plates (lexan/polycarbonate)\\ \\ \end{tabular} \\ \hline 
  References & \cite{Pedroni2011,Koschik2016} & \cite{Mizushima2017,Noda2017,Noda2016} & \cite{Vilches-Freixas2020}\\ \hline 
\end{tabular}
}}
\end{table}

\newpage
\section{Limitations and avenues for improvements}
\subsection{Accelerators}
\label{Section_Accelerator}
\begin{figure}[h!]
    \centering
    \includegraphics*[width=0.95\textwidth]{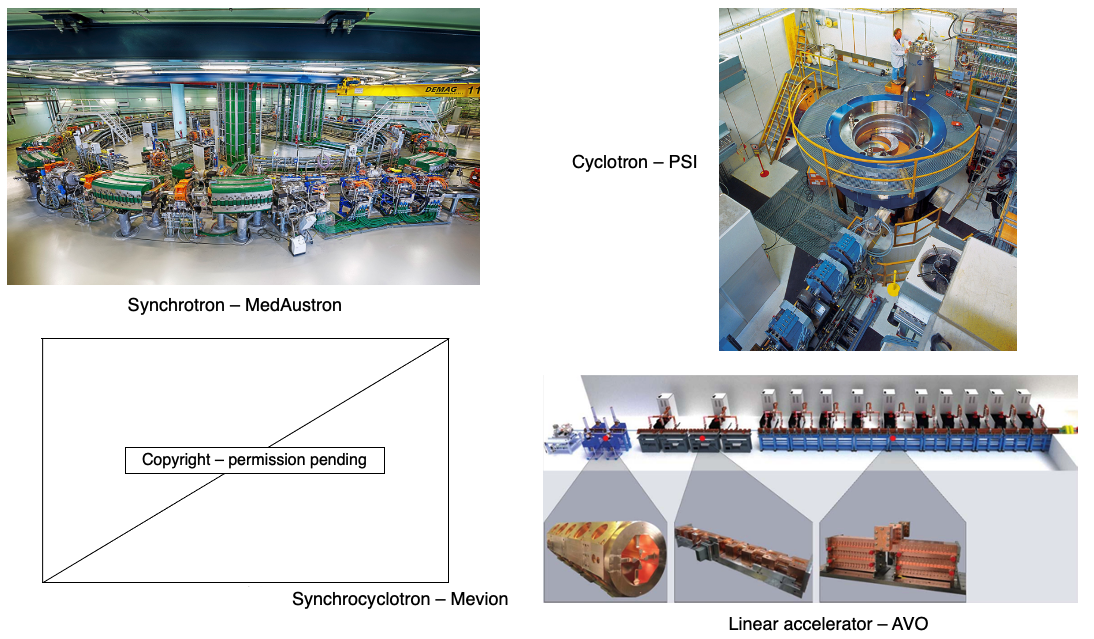}
    \caption{Synchrotron at the MedAustron facility \cite{Medaustron_image} and COMET cyclotron (for PBT) at PSI \cite{PSI_cometcyclo}. Gantry-mounted Mevion S250i synchrocyclotron \cite{Mevion_s250i}. Depiction of the AVO LIGHT LINAC with the radiofrequency quadrupole, side coupled drift tube linac and coupled cavity linac sections highlighted \cite{Ungaro2017}.}
    \label{F_Accelerators}
\end{figure}

The timing structure of delivered beams varies significantly between different types of accelerators, as they have different technological and safety limitations. In this section we present the main operational patterns and optimisation challenges for synchrotrons, cyclotrons, synchrocyclotrons and other accelerators that may soon be available for PBT or CPT facilities (Figure \ref{F_Accelerators}).

\subsubsection{Synchrotrons} 
The most common operation pattern for synchrotrons starts by injecting a beam pulse from a radiofrequency quadrupole (RFQ) and linear accelerator (LINAC) at up to $\sim$5 MeV/u (order of 10--100 \textmu s) of $\sim 10^{11} $ protons (or $\sim 10^{9} $ carbon ions) into the main synchrotron ring. The beam is accelerated in $\sim$1 s to the desired energy, then slowly but continuously extracted and transported to the irradiation room to treat a specific treatment slice. While extraction processes are typically capable to deliver the entire accumulated charge in $<$100 ms, the process is intentionally extended to match the maximum safely supported scanning speeds in the transverse plane (to order of seconds). After delivery to one IES is completed, the particles still circulating in the synchrotron are dumped and all the magnets of the main ring are ramped to their maximum current and back to injection levels in typically $\sim$1--2 s. This process, often called `\textit{magnet washing}', ensures reproducibility of the magnetic field in the presence of hysteresis of ferromagnetic components. If the number of particles injected into the synchrotron is not sufficient to deliver the required dose at the specific energy, more pulses will be injected, accelerated, and delivered. For injection pulses of the order of $\sim$ 10\textsuperscript{11} protons, only the treatment of large lesions ($>$1 L) in combination with hypofractionation requires more than one injection for some of the energy layer. The total time required to change energy (or refill the main ring) is then of the order of $\sim$2--4 s.\\

Two significant technological solutions have been implemented so far which drastically reduce the time required to change energy/refill the ring. The first is to reduce the time required to wash the magnets to a few hundreds of ms by employing an active regulation of the magnet power supply output, based on live measurement of the magnetic field \cite{Feldmeier2012,Bressi2018}. The technology necessary for active regulation of quadrupoles or sextupoles has not been implemented in clinical machines yet and therefore they still require washing, taking $\sim$100's ms. The total time to change energy/refill is then reduced to $\sim$1--2 s. \\

The second technique is called multiple energy extraction (MEE) operation (or extended flattop operation) and aims to reducing only (but drastically) the time required to change energy. In MEE \cite{Iwata2010} the beam can be extracted across several energy levels in a single spill; this enables the delivery of successive IES without needing to wait for re-/acceleration. The unused part of the beam circulating after completion of a slice is re-accelerated (or decelerated) instead of being dumped in preparation for a re-injection. Although the process is not completely lossless, it requires only roughly $\sim$100 ms \cite{Mizushima2014,Mizushima2017} to change beam energy. Younkin et al. \cite{Younkin2018} performed a study to quantify BDT savings with MEE implemented at a synchrotron PBT facility, finding an average of 35\% reduction in BDT. The ELST was reduced by around 90\% from a typical value of $\sim$2 s to 200 ms with MEE. Additional savings could also be achieved by improving charge, extraction limits and charge recapture rates: these depend on the performance and limits of the synchrotron. Note that in this scenario the re-injection times are not improved from the typical operation mode.\\

The extraction mechanism most often used in synchrotron ion beam therapy facilities is based around a slow resonant mechanism which is usually driven by a transverse excitation (RF knockout (RF-KO) \cite{Noda2002AdvancedRS,Krantz2018}) or a longitudinally slowly induced energy change (Betatron magnet \cite{Bressi2018,Badano1999,Pullia2016}). The former technique can keep the beam bunched throughout the extraction process, making re-acceleration (or deceleration) a delicate but feasible process. The latter requires the de-bunching of the beam, which makes further re-acceleration (or deceleration) theoretically possible (RF front acceleration), but extremely challenging, loss prone and potentially time consuming. Facilities not originally designed for MEE operation exhibit the trend to implement RF-KO first rather than attempting other retro-fitted techniques \cite{Savazzi2019,DeFranco2017,DeFranco2018}. The direction of energy change is typically fixed (e.g. always increases or decreases), because an up-down energy scanning would violate the reproducibility of the main ring magnetic field due to hysteresis effects. A magnetic field active regulation control would be necessary for this feature, but it has not been implemented for quadrupole and sextupoles magnets to date. \\

Alternative extraction techniques which are compatible with bunched beams are based on optics changes often used in larger synchrotrons for non-clinical applications \cite{Kain2019}. These could be applied in the future but so far promise limited advantages for small machines dedicated to therapy. The limited benefits offered by MEE operations for hypofractionated treatments of large lesions can be overcome by increasing the particle filling of the ring. Although the most pursued R$\&$D is a reduction of the facility footprint, LINACs capable of accelerating orders of magnitude larger currents of protons and light ions to $\sim$10 MeV exist or are being developed \cite{Garoby2021,Cara2016,Vretenar2020}. An extreme example is LIPAc \cite{KONDO2020111503} which has already demonstrated the acceleration of 5 MeV, 125 mA deuteron beam; this is $\sim$50 times more than the currents typically injected in synchrotrons for therapy.\\

Although it is not the focus of this work, it is worth mentioning that often synchrotrons are chosen by therapy centres also for their capability and ease of delivering multiple particle species (proton, He, C, O, etc.). Assuming multiple ion sources are used for each species. The time required to switch particle species is driven by the change and stabilisation of the field in the injector's and low/medium energy beam transportation's magnets. In some tests at NIRS-QST this was chosen to be $\sim$20 s \cite{Mizushima2019} but could potentially be reduced to only few seconds with dedicated design and development of the source, injector and overall control system \cite{Sokol2019}. 

\subsubsection{Cyclotrons \& Synchrocyclotrons}
The most common choice for proton therapy is the cyclotron (mostly isochronous cyclotrons), which always accelerate protons to a single (maximum) energy. The beam is typically available as a continuous wave (CW) beam with a micro-bunch structure of 100's MHz. As the extraction energy is fixed, the energy can only be changed by inserting material in the beamline to degrade the energy. This method produces large losses especially when selecting energy at the lower end of the spectrum ($\sim$90\% of the beam can be lost), creating a radioactive hot spot in the location of the system. The downstream beam has a very large distribution of energies, not suitable for precise 3D conformal dose delivery. Therefore, an energy selection system (ESS) usually follows the energy degrader. The ESS consists of several devices (degraders i.e. carbon or graphite wedges, collimators, slits, magnets, diagnostics etc.) which are necessary to modify the beam to have the correct parameters for treatment. The transmission, quality and distribution of the beam is affected by interactions with objects in the beam path (the increase in distal penumbra from the energy spread can be up to 10 mm) \cite{Farr2018}. The optics is designed to create a section with a large dispersion, where slits are inserted to trim the beam and reduce the energy spread. This also has implications for the gantry, discussed later.\\

Compared to synchrotrons, the time taken to mechanically insert the beam modifying devices is relatively quick ($\sim$10's ms) for small energy adjustments. The use of actuated static wedges with time compensation and fast deflecting magnets (range adaptation) is reported to be the fastest method, changing energies in less than 20 ms \cite{Chaudhri2010}. At PSI the wedge positioning takes 50 ms, as reported by Pedroni et al. \cite{Pedroni2011} and the fastest energy modulation times are achievable on gantry 2 at 80 ms. Delays are caused by stabilisation of the dipole magnets in the beam transport line and gantry. The direction of the energy change is not limited by the accelerator, but the reproducibility of the magnetic field in the ESS. The accelerator itself does not limit the direction swap of the energy change, which can change up and down in any sequence. The hysteresis of ferromagnetic components in the beam transport typically restricts fast energy change to one direction only, with a magnet wash required before each direction swap. The BDS magnets must be ramped to accommodate different energies and to preserve the beam position at isocentre: it is the time taken to vary and reset the magnetic fields in the BDS which determine the ELST time. This also holds true for synchrotron facilities however the extraction times far surpass these at present. Although designs for cyclotrons dedicated to particle therapy with He or C exist \cite{Jongen2010IBAJINR4M,Jongen2010REVIEWOC}, no facility has so far been developed yet.\\

Finally, we consider superconducting synchrocyclotrons, which produce a pulsed beam as they are not isochronous. This type of accelerator can bypass the limiting constraints of the BDS magnets as – at least for protons – it can be made into a single room system. The accelerator is gantry-mounted, requiring the entire machine to rotate around the patient. Therefore energy changes are performed using an energy modulation system; like a regular cyclotron this comprises polycarbonate plates, range shifters, absorbers or other devices which physically attenuate the beam \cite{Kang2020a,Zhao2016}. ELSTs as fast as $\sim$50 ms, for changes of 2.1 mm in water equivalent thickness \cite{Vilches-Freixas2020} have been achieved. Although these have a much smaller footprint and fast energy modulation, the achievable beam parameters and pulse structure are insufficient for continuous PBS delivery \cite{Paganetti2021}.\\

\subsubsection{LINACs \& FFAs}
\label{Section_OtherAcc}

A comparison of minimum baseline figures for each of the considered accelerator types are shown in Table \ref{T_AccBaselineFigs}. We discuss two further accelerators which are not yet in clinical usage, but have potential to overcome existing limitations. \\

Linear accelerators are already ubiquitous in hospitals as compact sources for conventional XRT. Using a LINAC for protons or ions is more challenging: they are physically much larger, in part because the velocity of the particles changes significantly for proton and ions in the clinical energy range, thus the physical length of accelerating gaps must accommodate for this change throughout the accelerator. A proposed LINAC-based solution from Advanced Oncotherapy \cite{physworld2018} includes: a high-frequency RF quadrupole design at 750 MHz, originating from CERN, and a side-coupled 3 GHz LINAC for the high-energy accelerating section originating from the TERA foundation \cite{PhysRevAccelBeams.20.040101,DeMartinis2012}. Above a threshold minimum energy ($\sim$70 MeV for protons), modular cavities are used to enable the LINAC to change energy. This is an unusual LINAC design, but enables the beam energy to be precisely regulated pulse-by-pulse at rates of around 200 Hz. This translates into a minimum time for energy change of just 5 ms \cite{Farr2018}. If lower ELSTs are required, technology capable of supporting kHz pulse rates exists. A LINAC is capable of switching energy in any direction but the 
bottleneck at present would be in the magnetic reproducibility in the BTL and gantry.\\

Two key R$\&$D goals being pursued are footprint reduction to democratise proton therapy and the introduction ion beam LINAC based therapy. LINAC designs could also be adjusted to accelerate different particle species, with particle species switching times limited by magnetic field changes and stabilisation in the ion source and low-energy section. Cavities with increasingly higher accelerating gradients are being designed and proposed, exploiting synergies with accelerators developed for high energy particle physics experiments. This trend could also contribute to shrinking of the injector stage of synchrotrons. \\

Furthermore, recent studies on clinical suitability for LINACs show that the production of a stable spot-size with energy can result in increased conformality particularly for deep tumours \cite{Farr2020}, and the ability to vary spot size on demand while delivering protons at FLASH dose rates could lead to LINACs having greater conformality and larger tumour volume capability compared to cyclotrons \cite{Kolano2020}.\\

An alternative future option is the Fixed Field Alternating Gradient (FFA) accelerator, which has static (in time) magnets but a beam orbit which spirals slightly outward with increasing beam energy. Unlike the cyclotron, the energy range of the FFA is in principle limitless, so heavier ions including carbon can also be accelerated to clinical energies. Note that FFA is a method of designing accelerator optics and can be applied also to beamlines and gantries. We will discuss accelerators first, then implications for the BDS and BDT.\\

FFAs have many potential applications as they can be compact in size, with fast acceleration and high beam current \cite{Sheehy}. The largest limitation is that this technology is less well-established than synchrotrons or cyclotrons: few accelerator physicists, engineers and component suppliers are familiar with this type of machine. However, they were first proposed back in the 1960s and have been developing rapidly in the last two decades. A number of FFAs have been constructed, the most relevant to this work are the two 150 MeV proton FFAs constructed in the early 2000s in Japan \cite{Uesugi2008} with 100 Hz repetition rates which have been the subject of detailed beam studies and characterisation \cite{Sheehy2016}. New designs for proton and ion therapy, including superconducting designs, have not yet been prototyped or constructed.\\

In general this type of accelerator produces a pulsed beam. This leads to a key advantage of the FFA for charged particle therapy: fast variable energy extraction, usually with single-turn extraction\footnote{Slow extraction may be possible, but has not been studied in detail.}. Extraction can occur at any time in the acceleration process, dictating the beam energy. A second advantage is that the FFA removes magnet ramping, overcoming hysteresis or magnet washing issues, so the pulse repetition rate of the FFA can be much higher than a synchrotron: a rate of 1 kHz was the goal of a 2010 design study PAMELA (Particle Accelerator for MEdicaL Applications) \cite{Peach2013}. This is vastly different to the few seconds ($<$1 Hz) for a regular synchrotron or 50--70 Hz for a rapid cycling synchrotron.\\

This rapid pulse rate has a remarkable feature of enabling pulse-by-pulse flexibility across the entire clinical energy range without limitation in energy step or direction, with a choice of particle species. This can not only reduce the ELST to the order of ms and improve the BDT, but this feature could even open up the possibility of interleaved treatment pulses (carbon or proton) and imaging (high energy proton) pulses delivered through the same BDS. \\

Removing restrictions in energy variation imposed by existing technologies could open up treatment options that are impossible at present. Yet taking full advantage of the rapid energy changes enabled by FFAs, LINACs or other machines with fast energy variation would require the BTL and/or gantry to be able to accept and deliver the extracted beam. A number of designs exist for future gantries, and implications are discussed in the next section.

\subsection{Beam Delivery Systems \& Gantries}
At currently treating facilities, the ELST typically approaches the order of seconds with a commercially available BDS, much longer than the baseline values reported in Table \ref{T_AccBaselineFigs}. While the transverse scanning magnets are relatively fast; it takes around a hundred times longer to move the beam the same distance longitudinally. The time delays to change the magnetic parameters of the beamline are currently an issue for cyclotron facilities; for synchrotrons, the use of MEE has been implemented in specific instances but is not yet universal \cite{Younkin2018}. Nonetheless, it is clear that the ELST is contingent on magnet ramping speeds and will be prohibitive, particularly when considering emerging developments in the field. In fact, the cost of accelerators is in general lower than the BDS so improvements in accelerators alone will not be sufficient for CPT to reach levels of XRT adoption \cite{Durante2016}. Accordingly, Myers et al. \cite{Myers2019} have advocated that progress with accelerators need to be matched by the BDS in order to accommodate for fast energy variation. Two possibilities are suggested: the use of superior magnets or alternatively, to increase the energy acceptance range. The first option would involve low inductance magnets that require large currents and therefore higher build and running costs. The second option has been considered frequently in literature \cite{Schippers2011a,Owen2013,Gerbershagen2016} and is becoming more of a possibility with the realisation of superconducting (SC) technology and advanced magnet designs.\\

The BDS contributes to a significant share of the capital and total cost for a CPT facility \cite{Loeffler2013,Durante2016} so future iterations must be designed to reduce operational and construction costs. The BDS must be able to transport the beam with high accuracy (sub mm precision), at different specific energies and deliver the correct dose distribution reproducibly. For systems which possess a gantry, the downstream section of the BDS comprises of series of magnets to bend the beam at multiple angles which must be able to transport the beam to isocentre with the required parameters (spot size, position and intensity etc.). Consequently, the gantry is a physically large and complex mechanical structure: this amounts to considerable costs associated with the weight, size, construction and operation. Most modern proton facilities have gantries in order to deliver PBS which achieves the highest quality of treatment. \\

For heavier ions, costs are much higher as the gantry must accommodate particles with higher beam rigidity and added physical constraints introduce greater probability of errors \cite{Strodl2008}. Currently there are only a few facilities which deliver carbon-ion beam therapy (CIBT) using a gantry: HIT, Heidelberg, Germany has a gantry which has a footprint of 6.5 m $\times$ 25 m (radius $\times$ length), weighing $\sim$670 t \cite{Fuchs2008, Schardt2010, Haberer2004} and HIMAC, QST, Japan has a SC gantry, 5.5 m $\times$ 13 m weighing $\sim$300 t \cite{Iwata2013, Iwata2018a}. A second generation, compact SC gantry with a smaller 4 m $\times$ 5.1 m footprint was also developed with Toshiba \cite{Iwata2018}. Heavy ion facilities which do not employ a gantry are limited to delivery with fixed beamlines. 

\vspace{4mm}
\begin{figure}[htb!]
    \centering
    \includegraphics*[width=\textwidth]{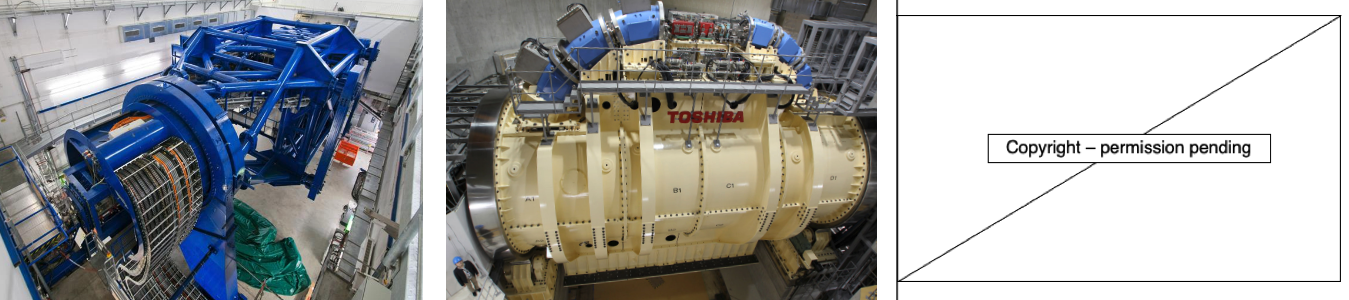}
    \caption{Gantry installations at HIT \cite{Pullia2016} and HIMAC, QST \cite{Iwata2018a} for CIBT. PSI PBT gantry 2 \cite{Owen2016}.}
    \label{F_Gantries}
\end{figure}

The use of SC magnets can dramatically decrease the weight and size of the gantry as higher fields (necessary for $>$1.8 T) can be achieved with comparatively fewer and smaller magnets. However, the costs for the magnets themselves and the operation of cooling systems may not be economical \cite{Gerbershagen2016}.\\

Furthermore, the question of the necessity of a gantry itself has now been raised: Flanz \& Paganetti et al. \cite{Paganetti2021} propose that the simplest way to reduce costs is to remove the gantry completely and in place have a fixed beamline. A study by Yan et al. \cite{Yan2016} indicates that for several disease sites (PBS head and neck cases), treatments could effectively be delivered gantry-less, requiring only a few fields with fixed geometries. There are other potential benefits to removing the gantry besides lower costs (maintenance, commissioning and also construction i.e. shielding) and the use of upright chairs is now being reconsidered. Seated positioning is typical for ocular treatments and for some specific disease sites; it was also historically the method used at pioneering facilities yet with less success than prone treatments \cite{Rahim2020}. Now, with the advent of modern delivery techniques, superior dose distributions can be achieved with seated treatments and clinical advantages with better immobilisation have also been reported \cite{Mazal2021}. This could be particularly effective for tumour sites which are difficult to treat due to motion and could also provide better patient comfort. A study by Sheng et al. \cite{Sheng2020} report that rotational and translational positioning with a 6D treatment chair is comparable in alignment precision and reproducibility to a standard robotic treatment couch. Another significant benefit is the prospect of enhanced integration with imaging; better conformity and registration between imaging modalities and the possibility of online imaging systems with the increased availability of physical space. However, full CT and MRI imaging would require these systems to be adapted to allow for seated patient positioning.

\subsubsection{Energy Acceptance Range}
\label{Section_EnergyAcceptance}
The energy acceptance -- or momentum acceptance -- is a limiting factor in existing beamlines and gantries. A typical momentum acceptance range is up to $\pm$1\% (approximately $\pm$2$\%$ energy acceptance) equating to changes of 5 mm in water equivalent depth, which is the usual spacing between each adjacent IES. This acceptance band is a technical limit corresponding to the maximum deviation from nominal beam momentum which can stably be transported by the magnetic beamline. Any such momentum deviation produces a change in trajectory (via dispersion) and the configuration of the magnetic elements determines the dynamics of the particle beam (beam optics) and stable range.\\

At present, for each IES change, the settings of all the BDS magnets must be changed synchronously whilst considering AC losses and hysteresis effects, requiring several checks and settling time for field stability \cite{Gerbershagen2016}. This maintains the correct beam parameters at isocentre and ramping typically occurs in one direction to reduce complexities. SC magnets with high ramp rates also experience issues with eddy currents, but their use in the BDS for heavier ions appears necessary in order to minimise size and weight. Increasing the momentum acceptance range enables the BDS to transport various beams with the same fixed magnet settings and therefore minimal dependence on their field ramping capabilities. \\

Several designs suited to protons have been proposed which use achromatic beam optics to suppress dispersion effects, reporting momentum acceptance ranges of $\pm$3$\%$ by Gerbershagen et al. \cite{Gerbershagen2016b}, $\pm$15$\%$ by Nesteruk et al. \cite{Nesteruk2019} and $\pm$25$\%$ by Wan et al. \cite{Wan2015}. For heavy ions, large acceptance can be achieved with new SC magnet designs (canted-cosine-theta combined function magnets) \cite{Durante2021}.\\

\begin{figure}[h!]
    \centering
    \includegraphics*[width=\textwidth]{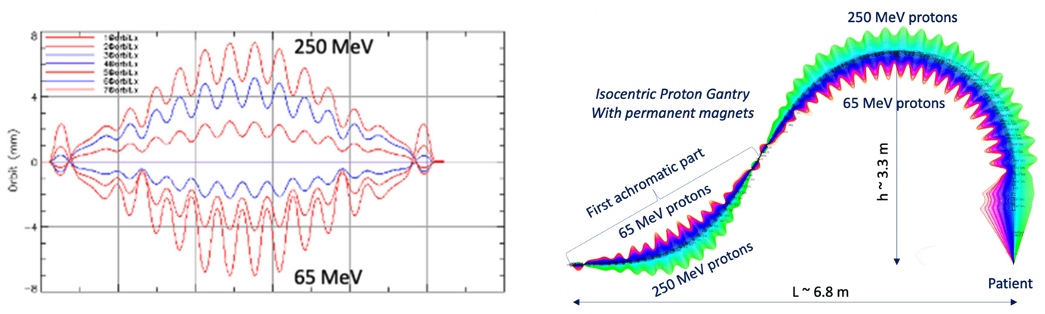}
    \caption{Orbit shape with varying energy (left) showing an energy acceptance range of 65--250 MeV. Orbit offsets within the permanent FFA gantry for PBT (right). Note the orbit offsets with energy are magnified for clarity in the right hand image and are around 15 mm, as shown on the left \cite{Trbojevic2021}.}
    \label{F_FFAgantry}
\end{figure}

Another optical configuration which enables a large energy acceptance (LEA) is the FFA concept (Section \ref{Section_OtherAcc}). With non-scaling FFA optics, combined function dipole and quadrupole magnets can be arranged in repeated cells in an alternating gradient configuration, resulting in strong focusing in both planes with small dispersion. This is stable for a wide range of energies and enables beam traversal along the beamline at multiple physical positions within the same fixed magnetic fields. Due to the low dispersion, small aperture magnets can be constructed, minimising size and construction costs. Multiple designs using FFA optics have been reported by Trbojevic et al. \cite{Trbojevic2007,Trbojevic2011} with a momentum acceptance range of approximately $\pm$20--30$\%$ using SC magnets for both PBT and CIBT; alternatively, novel Halbach type permanent magnets have been designed for a PBT gantry (Figure \ref{F_FFAgantry}) which accepts up to $\pm$35$\%$ \cite{Trbojevic2017,Trbojevic2021}. 

We note that alternative gantry designs to allow rapid beam delivery also include a novel method using high field SC magnets to produce a toroidal field capable of delivering beams from multiple directions in a fixed, steady state gantry \cite{Bottura2020a}. This removes time delays for change of gantry angle, but may have limitations in field size and challenges in achieving positional accuracy.\\

In contrast, for modern systems, the entire gantry rotates mechanically to deliver the beam from multiple entry angles. Therefore facilities which operate synchrotrons and employ slow resonant extraction produce fundamentally different beam distributions in the horizontal and vertical plane. This requires an optics matching stage in the transport line, often realised by inserting a thin foil that spreads the beam to equalise the particle distributions. The foil orientation might also require mechanical adjustment at different energies. Although typically this is not the bottleneck, it could potentially limit the ELST. Another arguably more elegant method is the use of a rotator \cite{Dorda2011} which physically rotates a set of magnets to match the optics for different angles, requiring no additional time for energy switching (apart from the magnetic field setting in the transport line). \\

Nevertheless, in general, increasing the energy acceptance of a BDS to enable a LEA suggests many benefits: these merits are mentioned among the many different design studies. There are both long term and immediate opportunities to improve upon existing technological capabilities. We now consider briefly a few aspects which pertain to future application of such a LEA BDS in clinical facilities. \\

The parameters of the magnets and the configuration of the optics design determine both the costs of the BDS and characteristics of the beam. This introduces a trade-off between the complexity and technical constraints imposed on the design and the achievable acceptance range: there must be an optimal range for which there will be maximal benefit. For example, Nesteruk et al. \cite{Nesteruk2019} describes that the $\pm$30$\%$ energy acceptance band can provide $\sim$70\% of patient treatments at PSI without the energy modulation requiring a setting change. The design and optimisation process will likely be driven by this requirement which will outline the cost benefit, particularly for the delivery of other particle types and heavier ions. \\

This also raises the question of the appropriate source-to-axis distance and positioning of scanning magnets either upstream or downstream \cite{Schippers2018}. What is clear is that in a novel BDS the parameters of the delivered beam must be clinically acceptable: energy spread (relates to range and beam penumbra), quality, size and shape (reproducible for every energy), position (spots must be positioned within precise margins) and also transmission (relates to particle rate for IES scans and current regulation with off-momentum particles). These properties must be consistent across the entire energy range and conform to performance and safety standards: rapid and accurate delivery cannot impinge on patient safety. Additional components (ripple filter, scattering foils etc.) may be necessary to moderate several beam characteristics upstream of the BDS \cite{Paganetti2021}.\\

The build of any BDS must be as robust as existing commercial systems (mechanically and operationally) and accommodate all the necessary components (beam diagnostics, nozzle, ESS etc.). For integration, the BDS needs to consider modularity for possible retrofitting or replacement of parts. Fundamentally, one must expect a lighter or smaller physical structure, also a simpler system in terms of functionality, servicing and tuning; these improvements along with cheaper running costs will further assist to lower overall expense.  \\

Finally, the adaptation of current control systems to manage a larger energy range is currently being explored. A recent study at PSI by Fattori et al. \cite{Fattori2020} demonstrates the clinical possibilities enabled by an increased momentum band to deliver PBS with real time tracking and enhanced rescanning capabilities. The prospect of energy meandering -- ramping beamline magnets bidirectionally (up and down) -- to further decrease the BDT is also presented. This combined with optimisation of the energy sequencing and layering offers higher flexibility and uptime in terms of the duty cycle \cite{Actis2018,Safai2012}. 

\subsection{Motion and Treatment Efficacy}
\label{Section_Motion}
The discussed benefits of decreased BDT have so far centred around the gain in delivery efficiency and therefore treatment efficiency or cost. Arguably however, the more compelling argument of a faster BDT is the potential of better treatment efficacy: treatment quality can be correlated to the efficiency of delivery \cite{Muller2016}. Future CPT facilities will need to be able to operate with shorter BDTs whilst ideally providing better quality treatments. As the BDT is dominated by the ELST, the accumulation of delays for each IES results in extended irradiation times; scanning sequences within 3--5 s or longer correspond with the respiration cycle and the effects of this motion are consequential for treatment \cite{Seco2009}. 

\subsubsection{Interplay Effects}
For PBS in CPT, there is an inherent challenge of utilising the BP due to uncertainties in the range and physiologic motion which compromise any dosimetric advantages. Heavy ions have regions of elevated LET and therefore greater sensitivity, this makes it more challenging to treat a wide range of indications, especially for moving tumours. The issue of motion during PBS delivery is twofold: both the target site and the beam deviate in position simultaneously, resulting in degraded dose distributions (interplay effects). 

\begin{figure}[htb]
    \centering
    \includegraphics*[width=\textwidth]{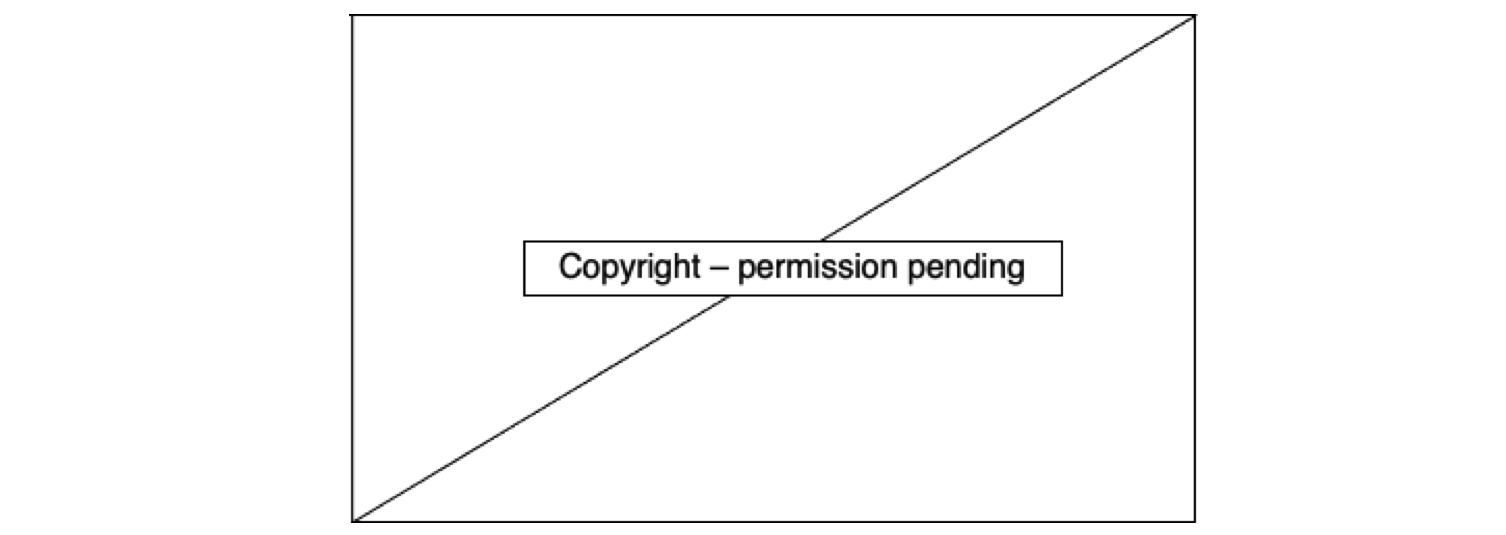}
    \caption{Delivery of a single IES with target motion (phases of movement are indicated by the plot and shown in blue, orange and green). The initially determined scan path in the target volume is shown in red. The raster scanned spots are translated outside the target due to motion which results in progressive degradation of the dose distribution \cite{Bert2008}.} 
    \label{F_InterplayEffects}
\end{figure}

Interplay effects (Figure \ref{F_InterplayEffects}) cause regional dose inhomogeneities due to under- and over-dosage, resulting in differences from the planned treatment distributions in each fraction. The clinical implications of interplay effects are well known \cite{Bortfeld2004,Lambert2005,Dowdell2013} and require a variety of motion mitigation strategies, many are commonly used in practice and more are also being developed. It is frequently recommended that a shorter BDT can decrease the extent of or even prevent interplay effects, if the BDS is capable of delivering the dose sufficiently fast. The overall length of treatment is important: shorter irradiation times are ideal to reduce the amount of intrafractional motion yet to correct for interplay effects, more fractions are beneficial. It may seem like these are in conflict as each occur at the detriment of the other however, the key factor is again the BDT itself: high dosimetric quality has been demonstrated to be achievable even with higher delivery efficiency \cite{Muller2016}. 

\subsubsection{Management \& Mitigation}
A shorter BDT is attainable by reducing the burden of long ELSTs: the longer the duration, the greater the need to minimise its impact. Work by Van De Water et al. \cite{VanDeWater2015} investigated the effect of a shorter BDT on plan quality by the direct reduction of ELSTs, using a self-developed method with their treatment planning system (TPS) which minimised the number of layers required to deliver a treatment with robust optimisation. It was shown that the BDT could be reduced by up to 40\% for a range of different disease sites without compromising treatment quality. There are also other methods to decrease the BDT including: increasing the IES spacing \cite{Kang2008}, varying the size of spots \cite{Grassberger2013}, using a range of non-uniform sizes \cite{Paganetti2021}, changing the dose grid size or spot spacing \cite{Hillbrand2010}, optimising spot sequencing \cite{Li2015}, scan path \cite{Kang2007}, or multiple criteria i.e. different weighted spots or resampling for selective placement of spots \cite{Muller2016,VanDeWater2013}. Cao et al. \cite{Cao2014} also present an energy layer optimisation method which increased the delivery efficiency whilst maintaining dosimetric quality. Each of these have varying effects on dosimetric metrics such as homogeneity, conformity indices or equivalent uniform dose. However, some associated benefits are not be quantifiable, such as patient comfort and further biological effects which may also contribute to better treatment outcomes. The purpose of any motion mitigation approach is to simultaneously maintain conformity and beam delivery duration \cite{Bert2008}. All of these corrective optimisation tools are designed to work around existing limitations in technology and if a new, faster BDS and accelerator system were made available, would either become obsolete, or could be made even more powerful to the benefit of both treatment efficiency and efficacy.\\

There are also a range of common techniques which have been translated from XRT to CPT, such as 4D planning and delivery \cite{Keall2004,Keall2006,Graeff2014}, a comprehensive overview is presented by Bertholet et al. \cite{Bertholet2019}. A simple method is to implement safety margins in treatment planning, expanding the clinical target volume to a planning target volume to account for uncertainties and dose delivery errors \cite{Albertini2011}. However, this has been demonstrated to be insufficient for more complex intensity modulated PBT plans \cite{Lomax2008a} and more robust methods are necessary to lessen adverse effects caused by the steep dose gradients and motion. Managing these is a highly complex task and there are a variety of motion mitigation strategies applied by different facilities, these are summarised in detail in \cite{Rietzel2010,Bert2011,Mori2018}. Some specific approaches include: breath-hold \cite{Gorgisyan2017}, beam tracking \cite{Saito2009}, gating \cite{Vedam2001} which can also be combined with rescanning \cite{Zhang2014,Schatti2014b}. The use of physical equipment to shape the beam has also been re-examined using ridge filters, 3D modulators \cite{Simeonov2017} and other beam shaping \cite{Moteabbed2016} or modulating devices \cite{Sanchez-Parcerisa2014}. Equivalent to passive scattering, the entire field can be delivered almost instantaneously which thus negates the effects of interplay \cite{Engelsman2013}. 

\subsubsection{Rescanning}
In addition to beam gating and tracking, rescanning -- also termed repainting \cite{Zenklusen2010} -- is a primary method used to mitigate intrafractional motion through repeated irradiation. In PBS, the small beams are particularly sensitive to motion and as this movement is generally periodic, dose errors can be statistically averaged out by increasing the number of fractions \cite{Phillips1992}. A minimum number of rescans must be performed for added benefit \cite{Grassberger2015a}, particularly for mobile sites such as the liver and lungs \cite{Dowdell2013}. Notably, the effect itself depends on the patient and beam parameters such as the direction, scan speed and path: characteristics determined by the accelerator and BDS. Bert et al. \cite{Bert2008} mention that by choosing favourable parameters, the severity of interplay effects can be lowered and quasi-eliminated if scan speeds are sufficiently quick. The significant concern with rescanning and other mitigation techniques is that they can extend treatment to unacceptable lengths of time. Even at facilities which offer fast dynamic energy modulation, the accumulation of BDT still surpass the time limits defined by the respiration cycle. The potential benefit of a faster BDS is the higher rescanning ability: this is dependent on the capabilities of the BDS, primarily its efficiency and the applied methods of delivery \cite{Klimpki2018}.\\

\begin{figure}[htb!]
    \centering
    \includegraphics*[width=\textwidth]{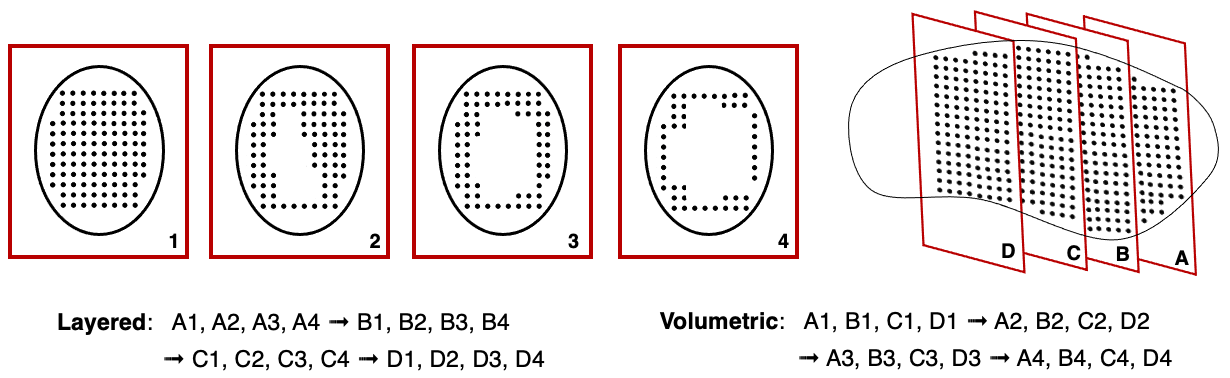}
    \caption{Possible IES pattern sequences for layered and volumetric rescanning.}
    \label{F_RescanningTechniques}
\end{figure}

\newpage
Another issue with rescanning is if the motion of the beam and patient are synchronised: this jeopardises the averaging effect. This can be avoided by ensuring delivery across the entire respiration cycle (i.e. phase controlled rescanning or breath-sampled rescanning) or introducing variations in the scan path by delays or randomness \cite{Bert2011}. There are several different patterns by which rescanning is performed (Figure \ref{F_RescanningTechniques}), most commonly it is done akin to typical delivery, by painting repeatedly across a IES before moving onto the next consecutive layer (layered rescanning). An alternative method is to move through the different layers first, returning to the same IES to paint subsequent distributions (volumetric rescanning). 

Volumetric rescanning is not employed clinically due to long ELSTs which make it impractical. Studies performed suggest several benefits as it enables additional scan paths and can alter the temporal correlation between beam and organ motion \cite{Rietzel2010}. Modifying the rescanning pattern to break the coherence of the beam structure with the periods of motion is an indicator of effectiveness; Bernatowicz et al. \cite{Bernatowicz2013} demonstrated in a comparative study that outcomes may be less dependent on when the irradiation occurs during the respiratory phase, if volumetric rescanning is performed. A study by Zenklusen et al. \cite{Zenklusen2010} suggests that combining volumetric rescanning with a fast delivery technique such as continuous line scanning can be an attractive method if it is possible to irradiate the entire volume within a single breath hold.

\section{Emerging Applications} 
\label{Section_FutureApplications}
The field of CPT is evolving rapidly and the limitations of even state-of-the-art technology are becoming apparent; the possibility of volumetric rescanning and other advanced techniques require the BDS to be able to deliver efficiently with fast energy modulation. A recent review by Mazal et al. \cite{Mazal2021} outline several of these proposed CPT approaches to reduce associated uncertainties, complexities and cost. We specifically examine technological constraints and discuss BDS improvements as relevant for FLASH and arc therapy. 

\subsection{FLASH}
The goal for treatment is to be able to irradiate the tumour sufficiently while sparing healthy tissue. This is represented by the therapeutic index (TI) and indicates the ratio between the probability of tumour control to normal tissue complication: improvements in delivery methodologies and treatment efficacy seek to increase the TI. There is always a trade-off with increasing the amount of dose delivered to the tumour, as normal tissue is simultaneously exposed to damaging radiation. Hyperfractionation and different approaches are commonly used in RT to vary the length of treatments to reduce toxicity and support the recovery of healthy tissue. Alternatively, some radioresistant tumours also respond well with hypofractionation. It is well established that the dose rate and irradiation time has an effect on cell response \cite{Hall1972,Hall1991} although it varies widely, dependent on biological parameters and the linear energy transfer (LET) related to the particle type \cite{Wozny2016}. For certain conditions, a minimal dose rate effect has been observed; this has prompted a surge of recent research activity to reconsider applicable irradiation time scales for better therapeutic outcomes. As such, the shift to ultra high `\textit{FLASH}' dose rates ($\geq$40 Gy/s in $\sim$100 ms) \cite{Favaudon2014} has gained significant interest and may have the potential to revolutionise RT. The promise of FLASH therapy has been reported from in-vivo studies which suggest an increased TI due to biological advantages by a reduction of normal tissue complications via the tissue sparing FLASH effect \cite{Durante2018}. However the specific mechanisms of this are complex and still yet to be clearly identified \cite{DeKruijff2020,Wilson2020,Hughes2020}. These drive the technical requirements necessary to induce the benefits and achieve clinical feasibility: the `\textit{beam parameter space}' determines the applicable radiation conditions such as the beam structure and particle type however much remains under investigation \cite{Esplen2020}. \\

It has been easier to modify existing clinical linacs to deliver FLASH with electron beams \cite{Lempart2019}. For ion beams there are difficulties with reaching the required dose rates, which demand an increase in beam current by several orders of magnitude for rapid irradiation of a clinically relevant volume. A number of CPT facilities have been able to modify their accelerators (mostly isochronous cyclotrons and synchrocyclotrons) for FLASH, producing proton beams and photon beams have been studied at large scale synchrotron research facilities \cite{Kim2021}. FLASH with heavier ions such as for CIRT is also being examined \cite{Zakaria2020}. \\

The necessary accelerator and beam delivery developments required to deliver FLASH with clinical protons are detailed extensively by Jolly et al. \cite{Jolly2020}. Alongside this is also the need for better instrumentation systems which can operate proficiently under FLASH conditions \cite{Nesteruk2021a}. Clinical FLASH trials have commenced \cite{Bourhis2019a} yet there is limited implementation due to many challenges: the technological requirements push the boundaries of and surpass current capabilities. Multiple studies \cite{Patriarca2018,Darafsheh2020} have developed experimental setups and investigated the applicability of these adaptations at clinical PBT facilities however, the optimal beam parameters (time structure, profile, range, uniformity, field size etc.) which will be feasible in practice, are still unclear. Furthermore, as ensuring precise beam delivery and positioning is difficult, the concept of `\textit{shoot-through}' FLASH \cite{Verhaegen2021} with protons is commonly performed where the use of the Bragg Peak may be considered redundant in place of maintaining a high, effective dose rate \cite{Diffenderfer2020}. \\

A fundamental challenge is achieving the requisite FLASH beam parameters for PBS delivery with clinical accelerator systems, along with safety and diagnostics restrictions \cite{VandeWater2019}. The generated beam intensity must be sufficiently high to realise the minimum effective FLASH dose rate and simultaneously, have adequate coverage and conformity over the applicable fields. A study by Zou et al. \cite{Zou2021} assessed the limitations experienced with cyclotrons by analysing the main machine parameters which influence delivery. The authors demonstrate that it is impossible to deliver FLASH dose rates across a planned 5 $\times$ 5 $\times$ 5 $\times$ cm\textsuperscript{3} SOBP region due to BDS dead times: magnet scanning speeds and significantly, the ELST. Nine IES scans were required and although applying a standard 1.5 s ELST was too slow, even the fastest clinical ELSTs of 50 ms and 80 ms, were also insufficient. For cyclotrons, the beam transmission and quality also suffers due to the presence of an ESS. A hybrid delivery scheme and modulation devices are suggested for reaching FLASH dose rates across the entire volume. Near instantaneous delivery should be targeted given the indicative 100 ms time frame necessary for the FLASH effect; this will also negate the effects of intrafractional motion. \\

\newpage
For conventional synchrotrons, one of the challenges is to be able to store enough particles in the main ring to deliver the entire portal, as the time required for a single re-injection and acceleration is typically considerably longer than 500 ms. Systems theoretically capable of injecting into the main ring at a suitable energy with a charge that exceed requirements exist \cite{KONDO2020111503}. However, developments in this direction are somehow orthogonal to the goal of footprint and cost reductions, as larger and more expensive equipment is generally necessary. When a large amount of particles is injected in the main ring at once, the interaction among the particles start to be less and less negligible. Although strategies exists to keep this space charge effect under control, the  effect is accentuated in smaller radius synchrotrons.

\subsection{Arc Therapy}
Similarly, the endeavour to speed up treatment times has recently renewed interest in arc therapy which is already a mainstream modality in XRT (i.e. volumetric modulated arc therapy). Radiation is delivered to the patient as the gantry rotates rather than with multiple fields of differently angled, fixed beams. It is possible to achieve higher quality plans with VMAT and a significantly faster BDT than with multiple static beams \cite{Holt2013}. This concept applied to PBT is termed proton arc therapy (PAT) \cite{Sandison1997}, combined with PBS, spot-scanning proton arc (SPArc) \cite{Ding2016} and with other ions (helium and carbon), spot-scanning hadron arc (SHarc) therapy \cite{Mein2021}. This delivery technique is highly complex and a fundamental challenge again lies with the capacity of the BDS: it must deliver reliably and continuously along rotational arcs, with the ability to switch quickly between energy layers \cite{Carabe-Fernandez2020}.  

\begin{figure}[h!]
    \centering
    \includegraphics*[width=0.9\textwidth]{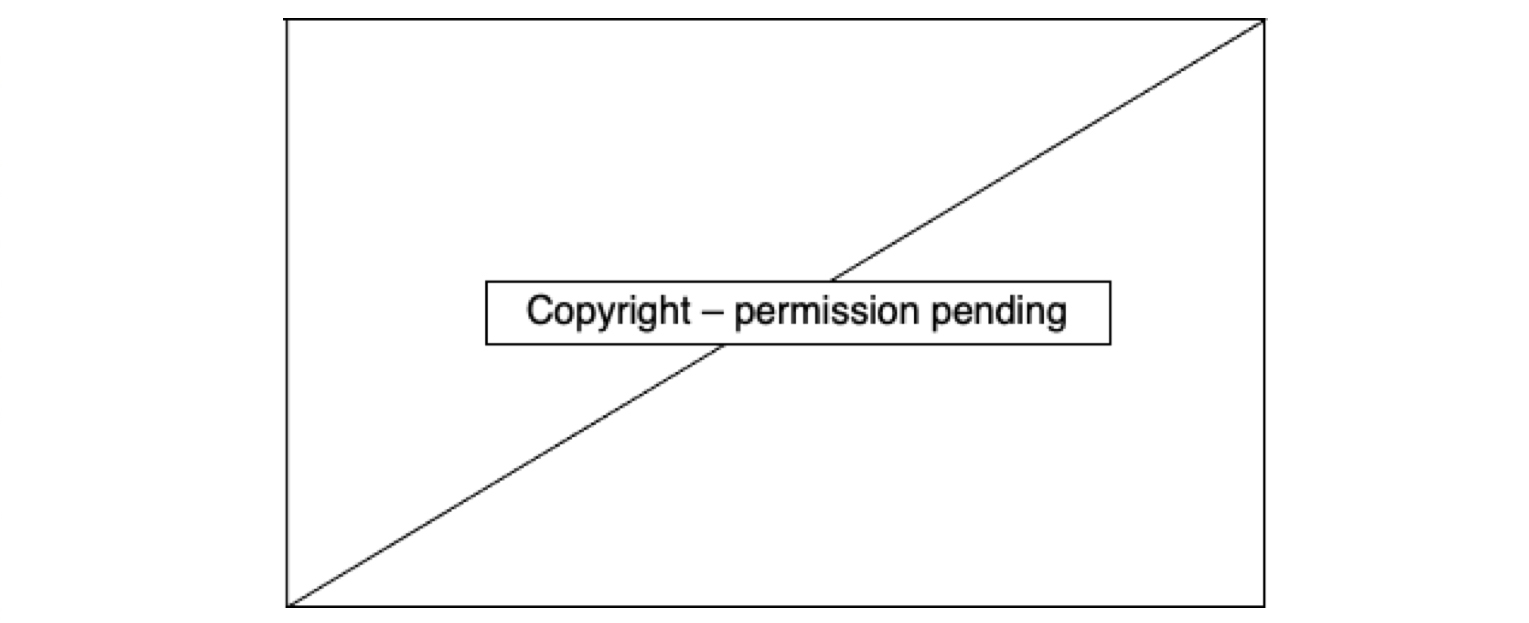}
    \caption{Dose distributions obtained from the delivery of mono-energetic PAT. Coronal view of 37 fields for a brain tumour treatment, applied by a couch rotation every 5$^{\circ}$ (left). Axial view of a tissue equivalent lung phantom using 35 fields, also 5$^{\circ}$ rotation (right) \cite{Sanchez-Parcerisa2016a}.}
    \label{F_ArcDelivery}
\end{figure}

In PAT (Figure \ref{F_ArcDelivery}), a spot scanned beam is delivered in a continuous arc which effectively dilutes the impact of range uncertainties, achieving a conformal dose distribution with also a reduction in standard entrance dose \cite{Seco2013b}. This allows greater flexibility in positioning high dose regions along the beam path and the potential for a much shorter BDT \cite{Schreuder2020}. Ding et al. \cite{Ding2018} have shown that good conformity and the possibility of a lower integral dose could be achieved with SParc however, treatment plans must be optimised for robustness and efficiency \cite{Sanchez-Parcerisa2016a,Gu2020}. Significant savings in BDT were reported \cite{Ding2016} with continuous arc delivery however this is not yet possible with current technology, due to complexities with gantry rotation and long ELSTs. As an alternative, conventional (step-and-shoot) spot delivery was suggested with a moving couch for fixed beamlines as well as better timing synchronisation. Carabe-Fernandez et al. \cite{Carabe-Fernandez2020} emphasised that although more investigation is needed, PAT has potential particularly for certain indications (brain tumours). Once again, in PAT the delivery efficiency is determined by the BDS and improvements are required such as better stability with beam current, positioning and notably, fast energy switching. The total BDT is a limiting factor \cite{Seco2013b} and treatment times can be reduced with shorter ELSTs which will also lessen dosimetric constraints due to a dependence on single fields. \\

Additionally, the possibility of an increased dose in the target volume with PAT may not translate to a higher degree of conformity however could exploit radiobiological advantages further by increasing the TI \cite{Bertolet2020a}. As with multi-ion radiotherapy (MIRT), combining different particle types for an effective mix of low- and high-LET regions could generate a higher dosimetric quality plan by utilising favourable characteristics. A planning study by Mein et al. \cite{Mein2021} evaluates SHarc with different field configurations using proton, helium and carbon ion beams (Figure \ref{F_MIRTdose}). The results demonstrate several possible clinical benefits such as a lower dose bath, minimisation of high-LET components on critical structures and better tumour control with normal tissue toxicity reduction. The use of multiple beam energies offer further gain over single or two-field plans however there are clear, unresolved technical hurdles with the present BDS and gantry systems which prevent the actualisation of this technique.

\section{Discussion and future directions}
\label{Section_FutureDirections}

It is contended that CPT will be a widely adopted modality once costs are comparable to XRT. Clinical outcomes will depend less on the capabilities of current technology and more on the physical characteristics of the delivered dose \cite{Flanz2013}. The CPT field is expanding in global prevalence and new and exciting advancements are on the horizon: delivery methodologies, design concepts and biological advantages. The current landscape encourages exploration of future long- and intermediate-term approaches which will allow CPT to exceed the ceiling achieved by state-of-the-art XRT \cite{Schreuder2020}. Evidently, next generation technologies and facilities are needed to address present challenges in CPT. \\

Several innovative concepts of delivery have emerged which offer hope of full exploitation of the unique advantages offered by CPT. However, among the multitude of prospective improvements for beam delivery, there is a common underlying goal: to address the challenges surrounding treatment time. This is a complex challenge as it encompasses more than the BDS and technology, as was examined in Section \ref{Section_BeamDelivery}. A shorter BDT results in not only a shorter treatment time, but is also consequential in terms of costs and treatment quality. The constraint imposed by the long ELST is a distinct hurdle in minimising the BDT: alleviating this would result in better treatment efficiency by reducing involved costs, increasing throughput and would also improve treatment efficacy.\\

Gantries account for the largest expense of facilities and more compact gantries have been designed for CPT however the general ambition has been mostly for size and cost reduction. Fundamentally, these hinge on the optical design and the parameters of the magnets, new possibilities are becoming feasible with SC technology however limitations and issues with ramping speeds still persist. Nevertheless, an alternative approach is to improve the beam transport capabilities of the BDS and redesign the optics to increase the overall momentum acceptance range; this has the potential to have a significant impact on treatment by eliminating the ELST dependency on technological bounds, thus shrinking the BDT. The feasibility of a LEA BDS as a solution to decrease the BDT is discussed in Section \ref{Section_EnergyAcceptance}. The prospect of a LEA BDS raises several challenges but has the possibility of achieving higher quality treatments at lower incurred costs, however it would need to be supplemented by further technological improvements throughout the whole system. Recent developments with accelerators have been pursued, these aspects and possibilities to decrease the BDT are discussed in Section \ref{Section_Accelerator}.\\  

\newpage
The ELST handicap on BDT is almost entirely dependent on technological limits: a myriad of different methods must be used in the clinic in order to provide effective treatments to circumvent existing capabilities. These add onto the delivery scheme and workflow; treatment planning optimisation is required or the BDS must be also be adapted to directly change the beam energy using mechanical components. This includes various approaches outside of the BDS such as patient specific devices as well as different accelerator feedback and extraction schemes. A shorter BDT has a significant clinical benefit and the impact of motion and interplay effects must be mitigated using various strategies, described in Section \ref{Section_Motion}.\\

\begin{figure}[h!]
    \centering
    \includegraphics*[width=0.9\textwidth]{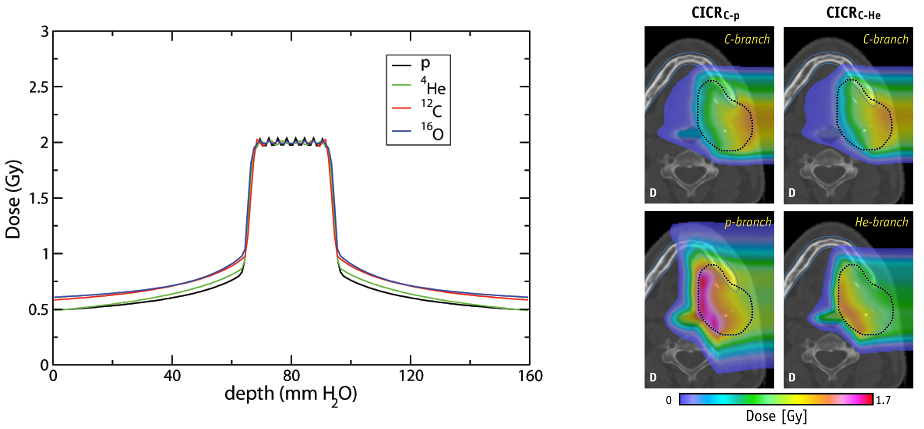}
    \caption{Conformal distributions can be produced by applying opposed fields using different particle types, optimising for physical dose (left) \cite{Tommasino2015b}. Combining beams of multiple particle types can generate distinctly different dose distributions, by positioning a chosen particle branch at the distal end of the field \cite{Kopp2020}. Images distributed under CC BY 4.0, BY-NC-ND.}
    \label{F_MIRTdose}
\end{figure}

A faster BDT and energy switching also drives developments toward a future BDS capable of delivering treatments for wider range of indications and with advanced CPT techniques. Several emerging applications are anticipated in the near future which will require an improved BDS for successful delivery such as FLASH and arc therapy. Several other anticipated developments in CPT are not discussed in detail but are also mentioned for context. Faster irradiation times go hand in hand with the need to ensure that treatments are still delivered with the necessary requirements of safety and precision. The importance of robust planning is also arguably higher for CPT than XRT but more challenging due to the aforementioned physical uncertainties, geometrical inhomogeneities and inter- and intra-fractional motion. This is again significant when considering different particle types; CIBT is expanding in addition to MIRT possibilities using beams of helium, oxygen, lithium etc. \cite{Tommasino2015b,Kirkby2020} which could further increase treatment efficacy. The combination of different ions offers a realm of new possibilities, by tailoring the desired LET and radiobiological attributes for different cancer types. Optimisation of these dose regions can offer more stable distributions and effective treatments (Figure \ref{F_MIRTdose}), such as using lower-LET particles for the sharper dose fall-off and higher-LET beams for hypoxic or radioresistant tumours \cite{Ebner2021}. 

MIRT is still a developing modality and technical limitations are primarily due to difficulties with the ion sources and long switching times. Considerations are also necessary with the acceleration and beam transport, as it is much more difficult to deliver beams of heavier ions given the associated beam dynamics, mechanical and physical requirements \cite{Yap2021}. The accelerator complex and BDS will need to be able to accommodate the range of different particle types; this may be selective based on characteristics such as mass and charge (i.e. mass-to-charge ratio $\leq$3 \cite{Volz2020}). For MIRT treatments, the beams are expected to be arranged such that the high dose regions fall appropriately within the target volume, hence this corresponds to different beam energies for each individual particle type. \\

The method of delivery must also be gauged, techniques such as minibeams and spatially fractionated RT are suggested \cite{Tinganelli2020}. Moreover, as the ion switching and ELST restrictions currently cause long BDTs, MIRT treatments have firstly been studied in a single field arrangement \cite{Kopp2020} and opposed fields \cite{Sokol2019}. The clinical flexibility and advantages are demonstrated however it is worth to note that technological possibilities may only allow sequential irradiation: this raises questions about throughput, quality assurance requirements, interplay effects, motion mitigation and fractionation schedules. The unknowns with the biological effects are also crucial, aside from the uncertainties with modelling and determination of treatment outcomes, the irradiation time structure and the division of the BDT between sources and fractions may introduce considerations with radiobiological chronicity \cite{Ebner2021}.\\

Simultaneous delivery with mixed beams is complex however has been performed for online monitoring and range verification with helium and carbon \cite{Mazzucconi2018,Volz2020}, exploiting the difference between BPs such that the carbon ions were used for treatment and helium for simultaneous imaging. Mixed fields are practically limited to synchrotron facilities as cyclotrons aren't able to achieve the acceleration requirements for heavier ions at clinical treatment depths, and the presence of an ESS changes the particle energy and velocity ratio \cite{Jolly2021}. Nonetheless, the strengths of MIRT and fundamentally CPT, can be achieved when limiting factors with delivery and motion are resolved; another important element with this is the need for precision imaging \cite{Schaub2020}.\\

Advanced imaging and techniques such as image-guided radiotherapy (IGRT) and adaptive radiotherapy (ART) are not yet clinically realised in CPT and the value of these have a higher potential for benefit in comparison with XRT \cite{Li2019}. Online imaging for treatment planning, monitoring and delivery is still at its beginnings, even with XRT \cite{Paganetti2021}. There are both technical and workflow challenges; it is an interventional process where plan adaptations should occur once certain changes are needed, triggering corrective measures. This is a resource intensive process, requiring both time and computational demands to re-plan and deliver the amended plan, and also online, in-vivo imaging and feedback systems. The ambition of real-time imaging has value in not just better treatment planning but can also enable continuous patient monitoring, as essential for online motion compensation and adaptation for 4D treatment delivery.\\ 

A promising avenue for this is with MRI guided PBT (MRPT), which has the capability of providing fast real-time imaging with superior soft-tissue contrast without the drawback of additional radiation exposure \cite{Hoffmann2020}. This approach is also still underdevelopment however there are several complex challenges with integrating MRI technology with a PBT system. The influence of the MRI magnetic field affects the trajectory of the proton beam, interfering with both the delivered dose distribution and resulting image quality. Corrections are required to compensate for the beam deflection and deviations in the treatment plan, dependent on the MR magnet field strength \cite{Moteabbed2014,Oborn2015}. The associated technological concerns relate to beam delivery and decoupling the PBS beamline from the MRI magnet, this requires an entirely new BDS design which can accommodate the physical and geometrical aspects of both systems \cite{Oborn2017,Schellhammer2018,Gantz2020}. The use of multi-modal approaches for advanced imaging is also of interest and combining i.e. CT with MRPT has exhibited benefits \cite{Bernatowicz2016,Paganetti2021}. 

\section{Concluding remarks}
The primary hurdle with CPT remains a question of cost, its availability and accessibility is still driven by the balance between cost and benefit: progressive improvements will contribute to decreasing the cost however future growth will depend on the extent of benefit \cite{Loeffler2013}. This is also influenced by various factors such as patient selection, clinical trials and scientific evidence. CPT still favours shorter treatments as it is difficult to immobilise patients for larger or complex lesions requiring extended treatment times (i.e. $>$30 mins). Increasing the range of accepted indications for treatment and capitalising on biological benefits (i.e. reducing fractions) supports the pursuit of reaching the same cost-effective levels as XRT \cite{Smith2017,Peeters2010}. \\

Furthermore, there are several challenges which impact the delivery efficiency and efficacy of treatments in CPT. We have reviewed the existing technical limitations and identified avenues for development in CPT. Focusing on the BDS, enhancements such as a LEA could reduce the limiting impact of the ELST on the BDT and shorten treatment times. This supports potential benefits such as cost reduction by expanding the utility of CPT and increasing the throughput of faster and higher quality treatments. Fast energy variation would also offer the capability of delivering advanced methodologies such as volumetric rescanning, FLASH and arc therapy. Improvements in beam delivery and related technologies enable the possibility of a future with cheaper, faster, precise and more effective CPT treatments.

\section*{Conflict of Interest Statement}
The authors declare that the research was conducted in the absence of any commercial or financial relationships that could be construed as a potential conflict of interest.

\section*{Author Contributions}
JY and SS: conception and design of the study. JY: literature review, manuscript preparation, writing, editing and figures. AD: writing and review. SS: writing, review and editing. All authors contributed to the manuscript and approved the submitted version.

\section*{Funding}
This work was supported by the William Stone Trust Fund and the Laby Foundation. 

\section*{Acknowledgements}
SS would also like to acknowledge the generous support of the Baker Foundation and ANSTO.

\bibliography{references}

\end{document}